\documentclass{article}

\usepackage{PRIMEarxiv}

\usepackage[utf8]{inputenc} 
\usepackage[T1]{fontenc}    
\usepackage{hyperref}       
\usepackage{url}            
\usepackage{booktabs}       
\usepackage{amsfonts}       
\usepackage{nicefrac}       
\usepackage{microtype}      
\usepackage{lipsum}
\usepackage{fancyhdr}       
\usepackage{graphicx}       
\usepackage{natbib}
\graphicspath{{media/}}     
\usepackage{amsmath}
\usepackage{graphicx}
\usepackage{enumerate}
\usepackage{natbib}
\usepackage{subcaption}
\title{Multiphasic stochastic epidemic models}

\author{
  Petros Barmpounakis and Nikolaos Demiris \\
  Department of Statistics\\
  Athens University of Economics and Business\\  
  Greece \\
  \texttt{\{barmpounakis, nikos\}@aueb.gr}\\
}

\begin{document}
\maketitle

\begin{abstract}
At the onset of the Covid-19 pandemic, a number of non-pharmaceutical interventions have been implemented in order to reduce transmission, thus leading to multiple phases of transmission. The disease reproduction number $R_t$, a way of quantifying transmissibility, has been a key part in assessing the impact of such interventions. We discuss the distinct types of transmission models used and how they are linked. We consider a hierarchical stochastic epidemic model with piece-wise constant $R_t$, appropriate for modelling the distinct phases of the epidemic and quantifying the true disease magnitude. The location and scale of $R_t$ changes are inferred directly from data while the number of transmissibility phases is allowed to vary. We determine the model complexity via appropriate Poisson point process and Dirichlet process-type modelling components. The models are evaluated using synthetic data sets and the methods are applied to freely available data from California and New York states as well as the United Kingdom and Greece. We estimate the true infected cases and the corresponding $R_t$, among other quantities, and independently validate the proposed approach using a large seroprevalence study.
\end{abstract}


The emergence on early 2020 of Covid-19, an infectious disease caused by the virus SARS-CoV2, has placed health systems around the globe under immense pressure. On March 2020, the World Health Organization declared Covid-19 as a global pandemic, and as of the end of September 2022 more than 6.5 million have died due to illness or complications of it. At the beginning of the pandemic in the absence of available vaccines or suitable medication the majority of governments around the globe resorted to Non-Pharmaceutical-Interventions (NPIs) in an attempt to stop the exponential spreading of the virus and reduce transmissibility. Such NPIs involved measures like work-from-home policies, school and university closures, stay-at-home guidance for people in high-risk groups and full lockdowns. 

These measures had an effect on reducing the transmissibility and resulted in spreading trajectories that could not be properly described by the standard epidemic models due to the resulting multiphasic nature of transmission. The first systematic technique to assess these interventions was due to \citet{Flaxman2020} who proposed a renewal equation model whose infection dynamics were modelled through a multilevel framework incorporating NPIs. We amend this model by inferring the points in time that the transmissibility changes as well as the magnitude of infectiousness in a data-driven manner. We determine the model complexity by using appropriate stochastic processes based upon variations of the Poisson process (PP) and Dirichlet process (DP)-based priors via their stick-breaking constructions (\citet*{MillerJASA}; \citet*{stickDP}).

Several models have been proposed in the literature for the estimation of multiphasic infectious diseases, particularly Covid-19. Briefly, a stochastic Susceptible-Exposed-Infectious-Removed (SEIR) model with a regression framework for the effect of the NPIs on transmissibility is used in \citet{key_epi_drivers_impact} while \citet{realtime_nowcasting}, \citet{Li2021-ur} and \citet{Chatzilena} use stochastic SEIR models where the transmission mechanism is described by a system of non-linear ordinary differential equations and the transmission rate is modelled by a diffusion process. Modelling the transmission rate as a random walk facilitates gradual and smooth changes in time. A piecewise linear quantile trend model was proposed by \citet{78d6270044e14bef8d2e1b9f1e7c8cbd}, a kernel-based SIR model distinguishing the different phases of the transmissibility in space was developed by \citet{kernel_SIR_Covi19} while \citet{Wistuba2022} incorporated splines to estimate the reproduction number in Germany.

Simpler forms of deterministic and stochastic multiphasic epidemic models have been considered before. In the context of modelling SARS-CoV2 transmission \citet{Flaxman2020} used an approach with fixed number, location and scale of the $R_t$ change. Related work based upon variations of Dirichlet process mixtures is presented in \citet{Hu2021} and \citet{CRESWELL2023}. In the former, the authors used a Mixture of finite mixtures (MFM) model on a Susceptible-Infected-Recovered-Susceptible model, while in the latter the authors used a suitably modified Pitman-Yor process but only for the scenario of fitting to the observed cases, thus dispensing with the effort to estimate the complete epidemic burden and the suitable adjustment for the reproduction number. The main advantage of the proposed methodology is the intuitive characterization of the epidemic in terms of multiple phases of transmissibility. The number and magnitude of the distinct phases are determined purely by data without explicitly using information about policy changes and NPIs. This approach should be central to a retrospective assessment of the NPIs: an evidence-based method for estimating the timing and effect of those interventions, minimising the risk of introducing several types of bias.

The paper is organized as follows. In section \ref{sec:mdlframework} we define the proposed compartmental process, elucidate its equivalence with renewal process-based models and describe the observation regimes of the data. In section \ref{sec:num_epi_phases} we complete the model definition by characterising the complexity regimes. Section \ref{sec:sim} assesses the proposed models via simulation experiments while section \ref{sec:Realdata} contains the application to data from California and New York state, the United Kingdom and Greece. The paper concludes with discussion.

\section{Modelling Disease Transmission}
\label{sec:mdlframework}

\subsection{Model Definition and Related Characterisations}
\label{sec:def_mdl}

The methodology for modelling the time-varying disease transmissibility has been implemented under two distinct but equivalent models, the compartmental Susceptible-Infectious-Removed (SIR) model and the seemingly simpler time-since-infection model with population susceptibility reduction. Here we define both models and delineate their equivalence. 

The model assumes that the population has size $n$, is closed (demographic changes during the course of the epidemic are ignored) homogeneous and homogeneously mixing. In the stochastic SIR model, an infected individual makes contact with any other individual on day $t$ at the points of a time-homogeneous Poisson process with time-varying intensity $\frac{\lambda_t}{n}$. This scaling is commonly adopted as it makes the contact process independent of the size of the population \citep*[e.g.,][]{Andersson2000-kf}. If these (close) contacts of an infected individual occur with a susceptible they result in an infection. Each individual remains infectious for a random time period $Y$. All Poisson processes in this construction are assumed to be independent. The disease reproduction number is defined as $R_t = \lambda_t*E[Y]$, $t=1,\dots,T$ where $T$ is the time horizon of the study.

For this model the expected number of new infections $c_{t+1}$ at day $t+1$ is given by: 
\begin{align}\begin{split}\label{eq:1}
    E[c_{t+1}] & =  S_t*\frac{\lambda_t}{n}*I_t*\Delta_{t+1-t}, \\
\end{split}\end{align}
with $I_t$ denoting the active set of infectives:
\begin{align}\begin{split}\label{eq:2}
    I_t & = \sum_{s=0}^{t} \sum_{j=1}^{c_s} P(Y_j > t-s) \\
\end{split}\end{align}
and $P(Y_j > t-s)$ the probability that individual $j$ infected on day $s$ remains infectious on day $t$. This probability is implicitly determined by the disease characteristics. Then (\ref{eq:1}) can be rewritten as 
\begin{align}\begin{split}\label{eq:3}
    E[c_{t+1}] & =  S_t*\frac{R_t}{n}*\frac{\sum_{s=0}^{t} \sum_{j=1}^{c_s} P(Y_j > t-s)}{E[Y]} = \frac{S_t}{n}*R_t*\sum_{s=0}^{t} c_s *g_s(t), \\
\end{split}\end{align}
where $g_s(t)=\frac{P(Y>t-s) }{E[Y]}$ is called the generation interval which defines the time from infection of an individual until the first infection they generate, see for example \citet*{SVENSSON201581F}, \citet*{SVENSSON2007300} and \citet{ErlangSEIR}. Note that equation (\ref{eq:3}) is used in the commonly adopted technique of \citet{Cori2013-hr} for estimating the instantaneous reproduction number. In that approach, the term $\frac{S_t}{n}$ which accounts for the depletion of the susceptible population is ignored since the aim is somewhat different.

One should also consider potential ‘superspreading’ events when certain individuals infected unusually large numbers of secondary cases (\citealp{Shen2004-fl}; \citealp{doi:10.1126/science.1086616}). We account for this variability assuming that the individual reproduction number is gamma distributed with mean $R_t$ and dispersion parameter $k$, yielding $c_t \sim Negative Binomial(E[c_t], k)$ \citep{Lloyd-Smith2005}.

\subsubsection{The Disease Reproduction Number}
The reproduction number $R_t$ is of great practical interest as it is used to assess if the epidemic is growing or shrinking. Here we consider two distinct instances of reproduction number. The effective reproduction number $R_e(t)=S_t*R_t$ describes the expected number of secondary cases generated by an infectious individual. Then $R_e(t) > 1$ and $R_e(t) < 1$ indicate that the epidemic is growing or shrinking respectively and reducing $R_e(t)$ below unity is the typical target of public health authorities. In contrast, $R_t$ quantifies contacts that may not always result in new infections, due to mixing with the immune proportion of the population. Therefore, $R_t > 1$ does not necessarily mean that the epidemic is growing. A detailed discussion about reproduction numbers can be found in \citet{Pellis2022}.      

\subsection{Observation Regimes}
\label{sec:obj_reg}
We consider two distinct observation regimes, one where the observed number of cases corresponds to the total number of infections, explained below, and whence the total number of infections is indirectly estimated, outlined in \ref{sec:unobs_cases}.

\subsubsection{Observed Infections}
\label{sec:obs_cases}
The regime where the total number of infections are observed may be of interest in its own right but may also be used for certain transmissible diseases, for example in the analysis of influenza like illness data when seroprevalence study information is available. Epidemic models are attractive for analysing such data and are naturally defined in terms of infector-infectee pair and the timing of such events. In reality however this type of data is rarely available. Disease monitoring is based on the daily reported infections, which are known to be susceptible to multiple problems, including a time lag between the timing of infection and symptom onset or testing positive. 

In the case of Covid-19 a large proportion of the population experiences asymptomatic or mild disease \citep{Ward2021} leading to severe under-reporting. Inference about the reproduction number can be robust when the reported cases are used if depletion of the susceptible population is accounted for, or if the observed proportion of cases remains constant over time. One way to validate this assumption is by sequentially performing seroprevalence studies to estimate the true disease prevalence and the proportion of unreported incidences. However, regular such information was not available in most countries. In the following subsection, we describe an alternative approach that dispenses with the need for this assumption.

\subsubsection{Unobserved Cases}
\label{sec:unobs_cases}
The case where infections may not be directly observed has been studied in a different context by \citet{demiris2014}. In the case of the pandemic, it became immediately apparent that the observed number of infections only partially accounts for the complete epidemic burden. An alternative technique was proposed by \citet{Flaxman2020} where the true cases were estimated by back-calculating infections from the daily reported deaths which are likely less prone to under-reporting. This method has the additional advantage of yielding an estimate of $S_t$. We adopt this approach for the second level of our model and the daily deaths are linked with the true cases via:
\begin{align}\begin{split}\label{eq:4}
    d_t & \sim NegativeBinomial(E[d_t],k)\\
    E[d_t] & =  IFR*\sum_{i=0}^{t-1} c_t *\pi(i)\\
\end{split}\end{align}
Accurate estimates of the infection fatality ratio ($IFR$) and time-from-infection-to-death distribution ($\pi(i)$) are necessary for estimating incidence, treated here as a latent parameter. The $IFR$ and $\pi(i)$ parameters may be calculated independently from external data or in a single stage, leveraging additional evidence from seroprevalence studies as illustrated in \ref{sec:Realdata}. 

\section{Epidemic Complexity Determination}
\label{sec:num_epi_phases}
The number of phases may be treated as a fixed but unknown integer or as a random quantity to be modelled and estimated from data. We describe two such models in the following two subsections.

\subsection{Deterministic Number of Phases}
\label{sec:fixed_num_epi_phases}

For the models described above `model complexity' refers to the number of epidemic phases. In \citet{Flaxman2020} the number of phases was a-priori selected and the times when the reproduction number $R_t$ changed were also predefined. The locations of these points were informed by the NPIs implemented by each government leading to a piece-wise constant reproduction number $R_t$, effectively assuming immediate effect of those NPIs. We also consider that $R_t$ is a piece-wise constant function and we amend this transmission mechanism by inferring the location and magnitude of $R_t$ changes directly from the data. The number, $K$, of epidemic phases is investigated using models with different $K$ values and the best model is selected using the Watanabe–Akaike information criterion (WAIC) \citep{waic} and Leave-one-out cross-validation (LOO) \citep{Vehtari2017}. The model is defined as follows: 

\[ 
R_t  = \left\{
\begin{array}{ll}
        r_1, \quad  t \leq T_1 \\
        ...\\
       r_{j+1},\quad  T_j< t \leq T_{j+1}\\
       ...\\
      r_{K},\quad  T_{K-1} < t \leq T\\
\end{array} 
\right. 
\]
\begin{align}\begin{split}\label{eq:5}
    r_j & \sim  f(\cdot),\quad r_j \in (0,\infty),\quad j=1,...,K\\
    T_{i+1} & = T_i + e_i\\
        T_1 & \sim  \operatorname{Uniform} \left({3,T}\right)\\
    e_i & \sim \operatorname{Uniform} \left({0,100}\right),\quad i =1,...,K-1\\
\end{split}\end{align}

\subsection{Stochastic Number of Phases}
\label{sec:stoch_num_epi_phases}

\begin{figure*}[hbt!]
    \centering
      \includegraphics[width=0.95\linewidth]{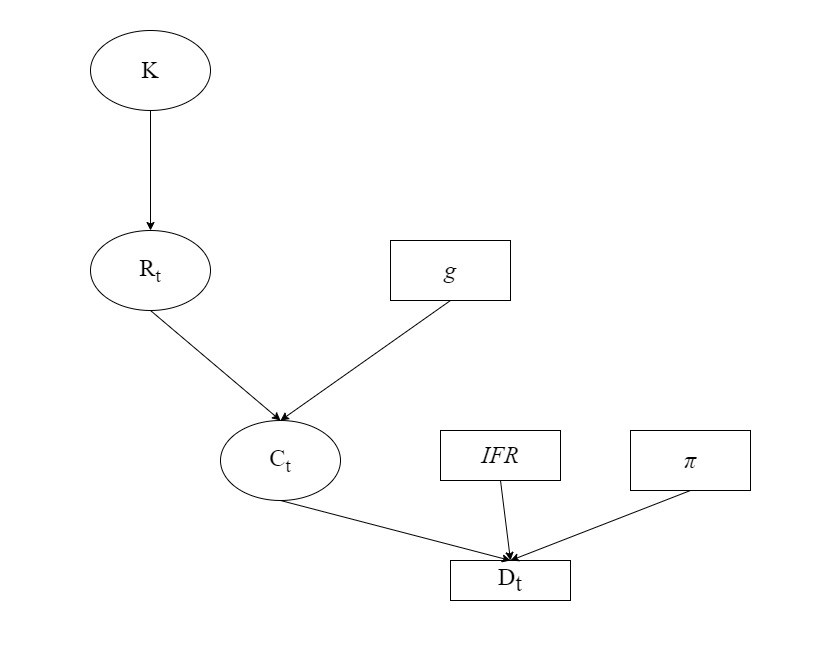}
    
\caption {Directed acyclic graph of the model. Ellipses denote parameters to be learned by the model. The number of phases K is estimated by the DP/PP model or via model selection criteria.
\label{fig:graph_model}}
\end{figure*}

Under the Bayesian paradigm, a natural but not trivial way is to treat the model complexity, here the number of epidemic phases K, as a parameter and learn its posterior distribution. The `reversible jump' algorithm \citep*[e.g.,][]{https://doi.org/10.1111/1467-9868.00095} could be used to explore the joint space of K and within-K models. Here we adopt a different approach and model $K$ as a characteristic of two stochastic models, the Poisson process (PP) and variations of the Dirichlet process (DP) \citep{Ferguson73}. For both processes, we use the stick-breaking representation, see \citet*{Miller2013-wh} and  \citet*{stickDP} for the PP and DP respectively, facilitating inference for $K$. The directed acyclic graph (Figure \ref{fig:graph_model}) represents the general structure of our modelling framework.

Estimating the number of phases of the epidemic and the associated location and magnitude of the $R_t$ changes can lead to identifiability problems for $R_t$ and its generative quantities, notably the total number of infections. In order to overcome such issues we explore both a single and a multi-stage modelling procedure \citep[e.g.,][]{Bhatt}. In the latter, at the first stage, the latent disease cases are estimated using a Gaussian Process (GP) model and then the medians of these latent cases are treated as data with likelihood given in (\ref{eq:3}). The GP for the estimation of cases is presented in the supplementary material.

\subsubsection{Poisson Point Process-based Model}
\label{sec:PoisPr}
We consider that the arrival of new phases in the time horizon (0,T] is driven by a time-homogeneous Poisson process with rate $\lambda$, with K growing linearly with time. Hence, following the first epidemic phase, the number, K-1, of new phases follows a Poisson distribution with rate $\lambda*T$ while the duration of each phase a-priori follows an Exponential distribution with rate $\lambda$. We follow \citet{Miller2013-wh} and use the representation:

\begin{align}\begin{split}\label{eq:6}
    R_{t} &= r_{z_t}\\
    r_j & \sim  f(\cdot),\quad r_j \in (0,\infty),\quad j=1,...,K\\
    z_t & \sim \operatorname{Categorical }\left({\pi_{1:K}}\right),\quad t =1,...,T\\
    \pi_K & = 1-\sum_{k=1}^K \pi_k,\quad K = min\{j: \sum_{i=1}^j T_i \geq T\} \\
    \pi_k & = \frac{T_k}{T},\quad k =1,..., K-1\\
    T_i & \sim \operatorname{Exponential}\left({\lambda}\right),\quad i =1,..., K_{max}\\
    \lambda & \sim \operatorname{Gamma}\left({0.02, 1}\right)\\
\end{split}\end{align}
truncating K at $Kmax= 100$, far higher than data-supported estimates.

\subsubsection{Dirichlet Process-based Model}
\label{sec:DirPr}
An alternative model for the number of phases is based on the DP and its stick-breaking construction:

\begin{align}\begin{split}\label{eq:7}
    R_{t} &= r_{z_t}\\
    r_j & \sim  f(\cdot),\quad r_j \in (0,\infty),\quad j=1,...,L\\
    z_t & \sim \operatorname{Categorical }\left({w_{1:L}}\right),\quad t =1,...,T\\
    w_{L} & = \prod_{k<L}(1 - v_k),\quad K = \sum_{k=1}^LI\{w_k\ge0\}\\
    w_l & = v_l*\prod_{j=1}^{l-1}(1-v_j),\quad l=2,3,...,L-1\\
    w_1 & = v_1, \quad v_i \sim \operatorname{Beta}\left({1,\theta}\right), \quad i =1,...,L-1\\\
    \theta & \sim \operatorname{Gamma}\left({1,1}\right)\\
\end{split}\end{align} 
where $L$ is the truncation point of the DP, set here to 36. Here K is increasing with the scaling parameter $\theta$.

\section{Simulation Experiments}
\label{sec:sim}
Simultaneously learning the parameters and the dimension of a model is typically a challenging statistical task. Here we adopt a simulation-based approach to inference whose details are given in the supplement. We assess the performance of our methods by simulating epidemics of various characteristics for 250 days. The epidemic model defined in (\ref{sec:mdlframework}) was used for simulating  daily infections and deaths. The population size was set at $10^8$ with $IFR=2\%$. The discretized infectious period and the infection-to-death interval are described in the supplementary material. The epidemic was simulated with 5 distinct increasing/decreasing phases resembling the observed Covid 19 outbreaks. The time-varying reproduction number was set as follows:
    \[ 
R_t= \left\{
\begin{array}{ll}
      1.5, & t \leq 60 \\
      0.95, & 60 < t \leq 100\\
      1.35, & 100< t \leq 150\\
      0.8, & 150 <t \leq 200\\
      1.8, & 200 <t \leq 250
\end{array} 
\right. 
\]

\begin{figure*}[hbt!]
    \centering
    \begin{subfigure}{0.49\linewidth}
      \centering
      \includegraphics[width=1\linewidth]{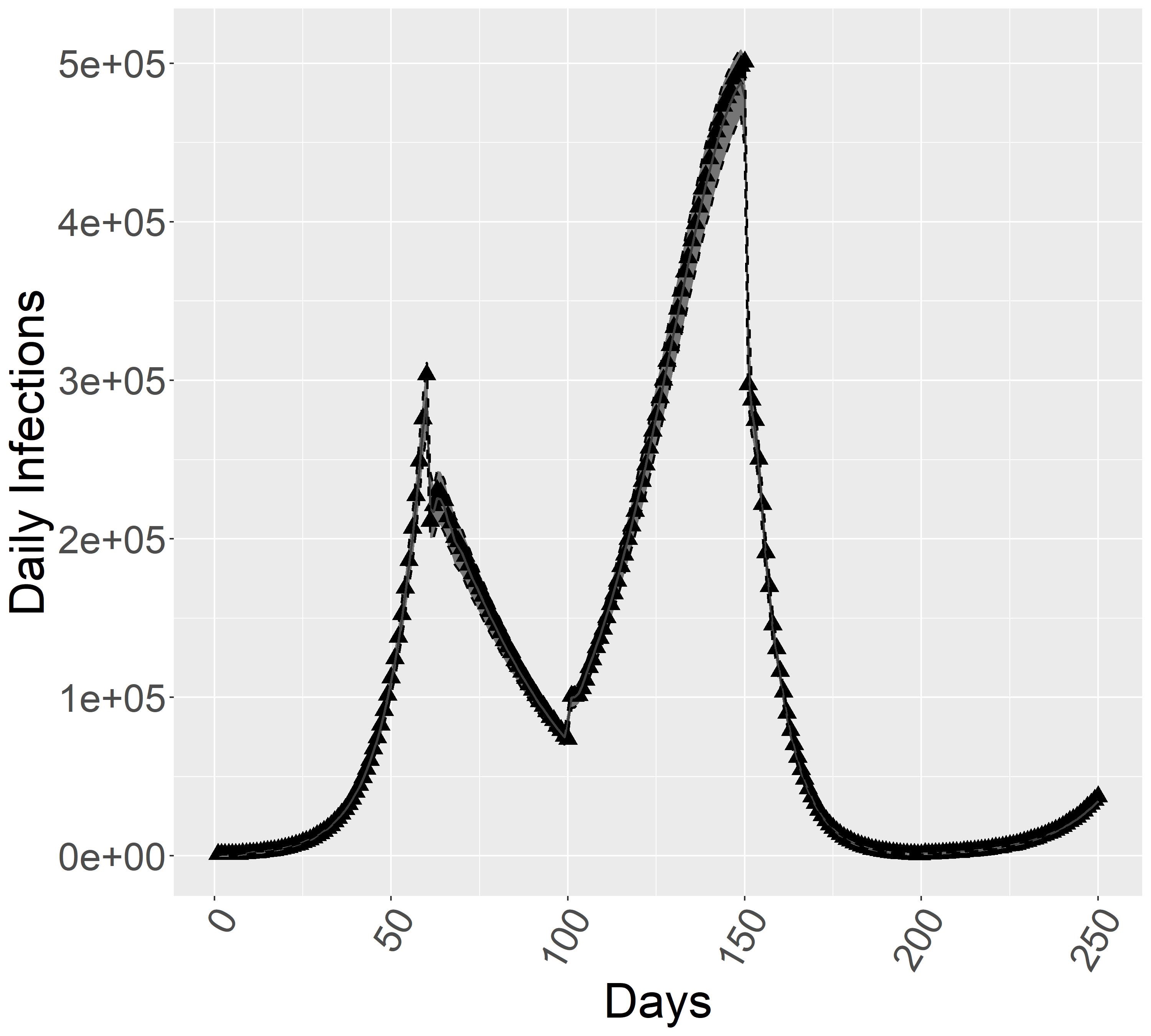}
      \caption{Simulated (triangles) and estimated daily infections with 95\% Cr.I. (line).}
    \end{subfigure}%
    \begin{subfigure}{0.49\linewidth}
      \centering
      \includegraphics[width=1\linewidth]{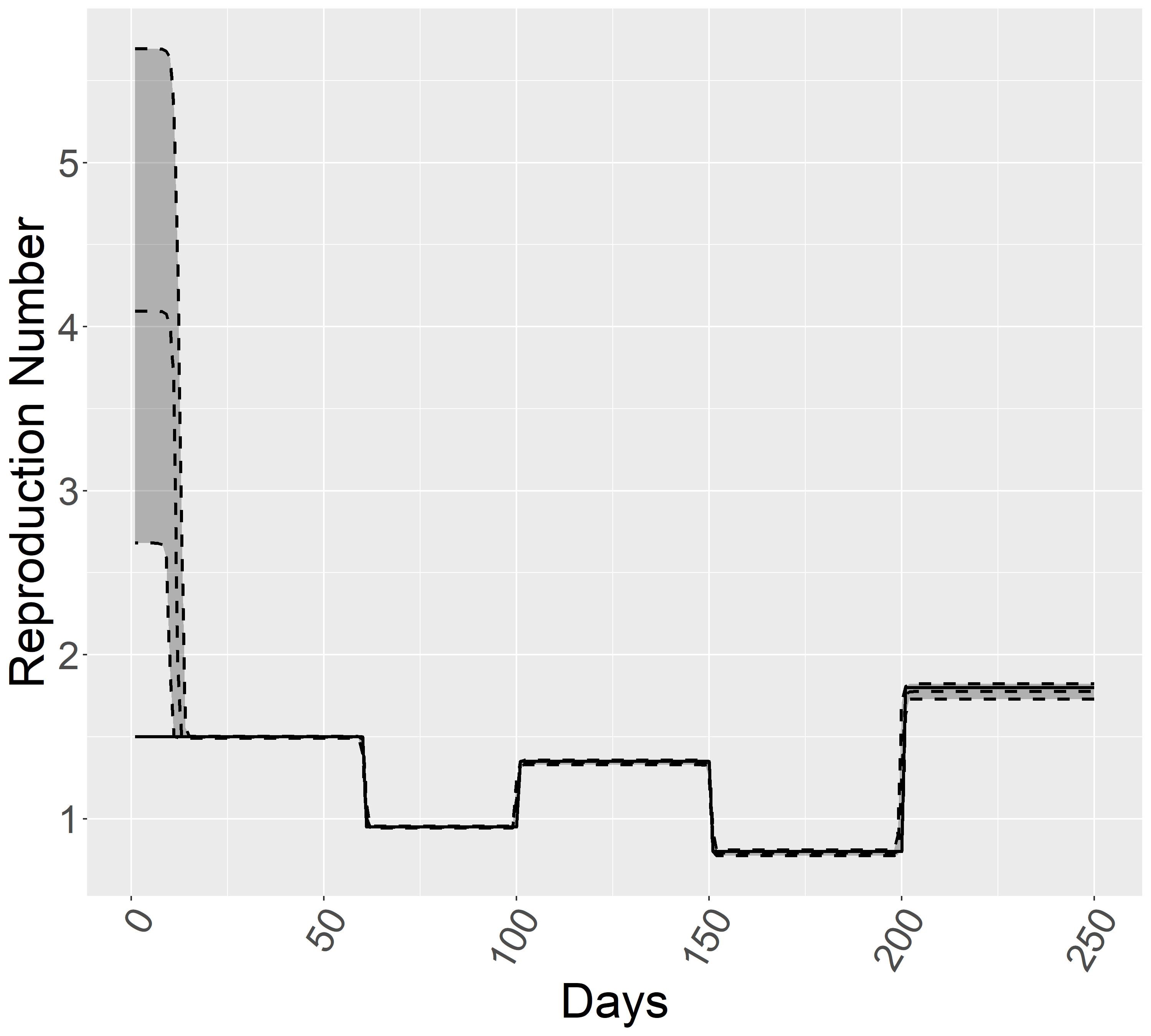}
      \caption{Real (solid line) and estimated reproduction number $R_t$ with 95\% Cr.I. (dashed line).} 
    \end{subfigure}
  \caption {Simulation and estimates based on observing deaths\label{fig:sim_5cp}}
\end{figure*}

Using the model in (\ref{eq:5}) and the daily deaths as data the lowest WAIC and LOO selected 5 changepoints. Models with varying (3, 4, 5 and 6) number of changepoints incorrectly identified the first 10 days of the simulation as a distinct phase (Figure \ref{fig:sim_5cp}). This can be attributed to the lack of information at the start, a common issue in epidemic models. Following this period the model with 5 changepoints correctly identifies the different epidemic phases, including their timing and magnitude of change. The total daily infections (Figure \ref{fig:sim_5cp}) are also accurately recovered. Inference was initiated the day that 10 cumulative deaths were observed. Plots for the other models may be found in the supplementary material.

\begin{figure*}[hbt!]
    \centering
    \begin{subfigure}{0.5\linewidth}
      \centering
      \includegraphics[width=1\linewidth]{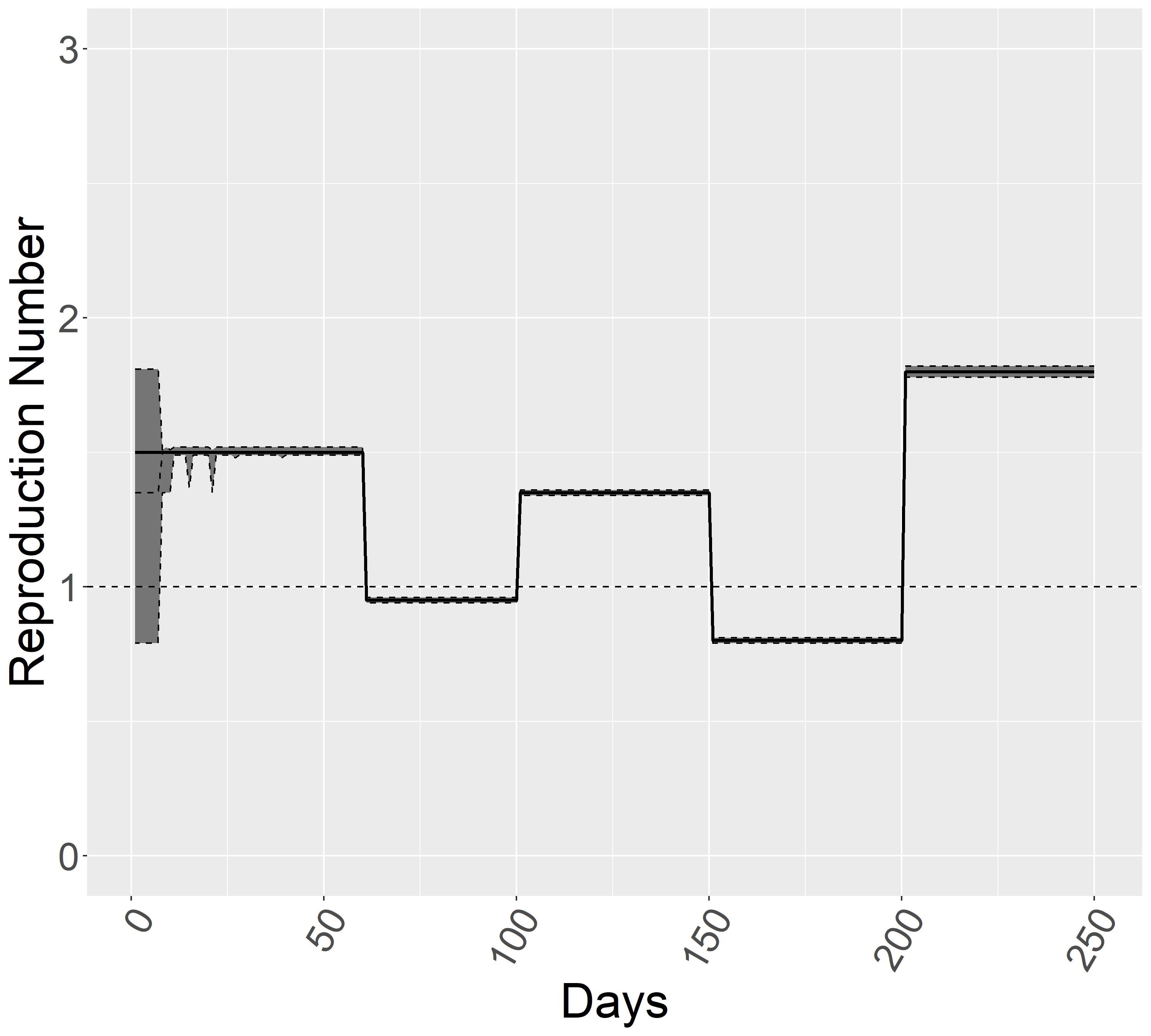}
      \caption{Dirichlet process model}
    \end{subfigure}%
    \begin{subfigure}{0.5\linewidth}
      \centering
      \includegraphics[width=1\linewidth]{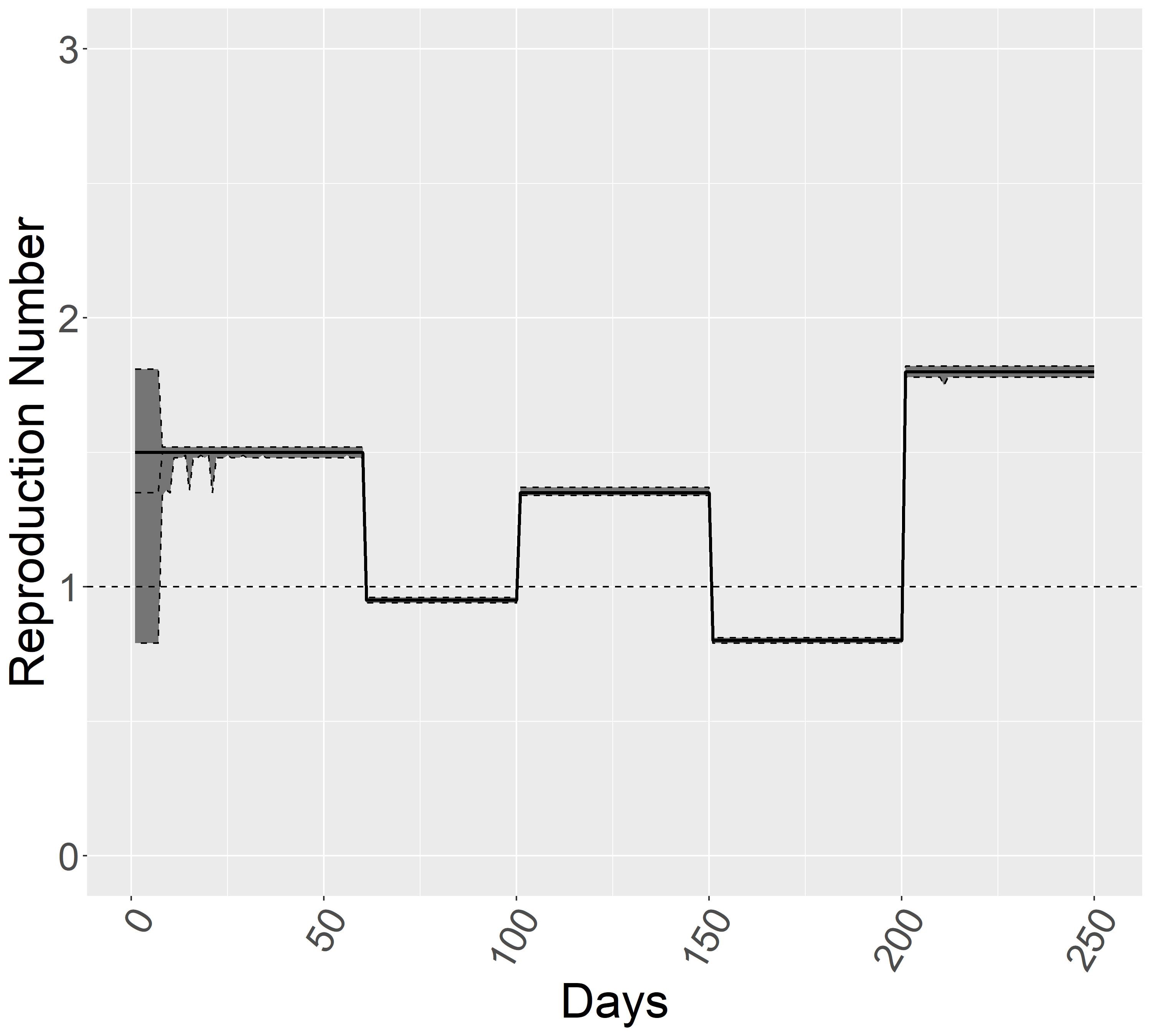}
      \caption{Poisson Process model}
    \end{subfigure}
    
  \caption {True (solid line) and estimated reproduction number $R_t$ with 95\% Cr.I. (dashed line) based on observing infections   \label{fig:sim_cases_DP_PP}}
\end{figure*}

In addition to the findings that the models correctly select the right complexity, it is interesting to summarise the model behaviour when investigating model misspecification. Broadly, these findings may be summarised as follows; when we fix the number of phases to be smaller than the true one then the model is correctly recovering the early ones while it is averaging the final phases leading to poorly fitted models. In contrast, when fixing $K$ to be larger than the true one then we essentially recover the true patterns and get a good fit. Hence, slightly overestimating model complexity is not materially affecting the recovery of the true signal. A list of detailed results is outlined in the supplement.

When fitting the models with a stochastic number of phases to daily infections, both the PP and DP models are precisely estimating the number of epidemic phases, the time of change and the true $R_t$ value (Figure \ref{fig:sim_cases_DP_PP}). The model was run for 100000 iterations and 8 chains. The analysis based on observing deaths is included in the supplementary material. Briefly, the intermediate phases of the epidemic are well estimated while the first and final phases are recovered with noise. The level of smoothing introduced by the cubic spline affects the noisy estimation of the cases; the lower the degrees of freedom the smoother the estimation of cases and subsequently the reproduction number.

\section{Real-data Application}
\label{sec:Realdata}

\subsection{Data Description and Prepocessing}
The models were fitted to daily reported deaths from two US states, California and New York and two European countries, the United Kingdom and Greece. The data are accessible from John Hopkins University and ECDC and the time horizon ran to the end of June 2021 when many NPIs were lifted. Due to a lack of data availability, the model does not account for reinfections. The age-standardized $IFR$ for each country was informed by the meta-analysis from \citet{COVID-19_Forecasting_Team2022-tv} accounting for time, geography and population characteristics. We allowed the $IFR$ to vary over time, accounting for the age structure of those infected, the burden of health systems and amendments in treating the disease. The infection-to-death time and generation interval were given a Gamma distribution with (mean, standard deviation) set to (19,  8.5) and (6.5, 4.4) days respectively.

\subsection{Analyses and Results}

\begin{figure*}[hbt!]
    \centering
    \begin{subfigure}{0.5\linewidth}
      \centering
      \includegraphics[width=1\linewidth]{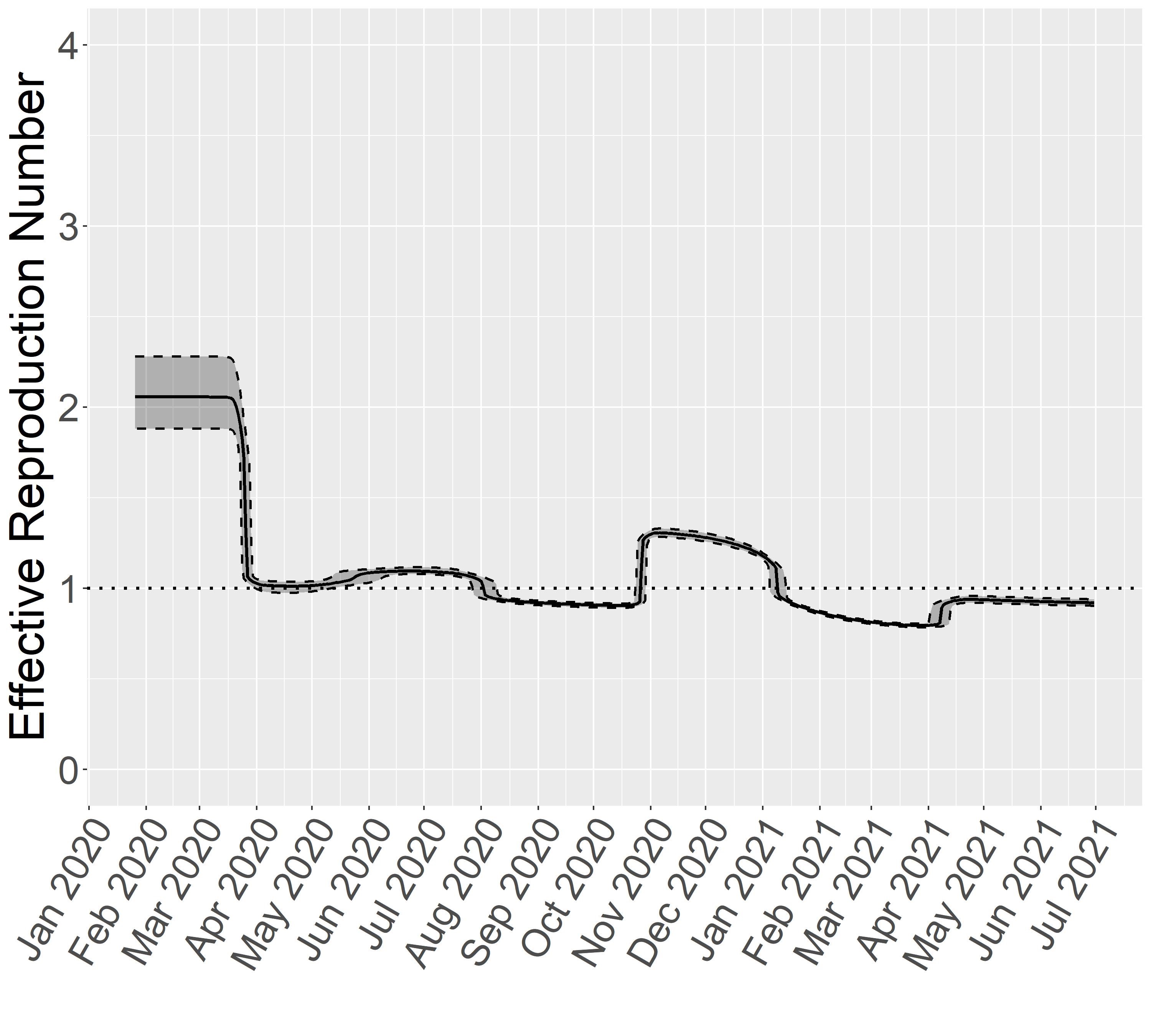}
      \caption{California state}
    \end{subfigure}%
    \begin{subfigure}{0.5\linewidth}
      \centering
      \includegraphics[width=1\linewidth]{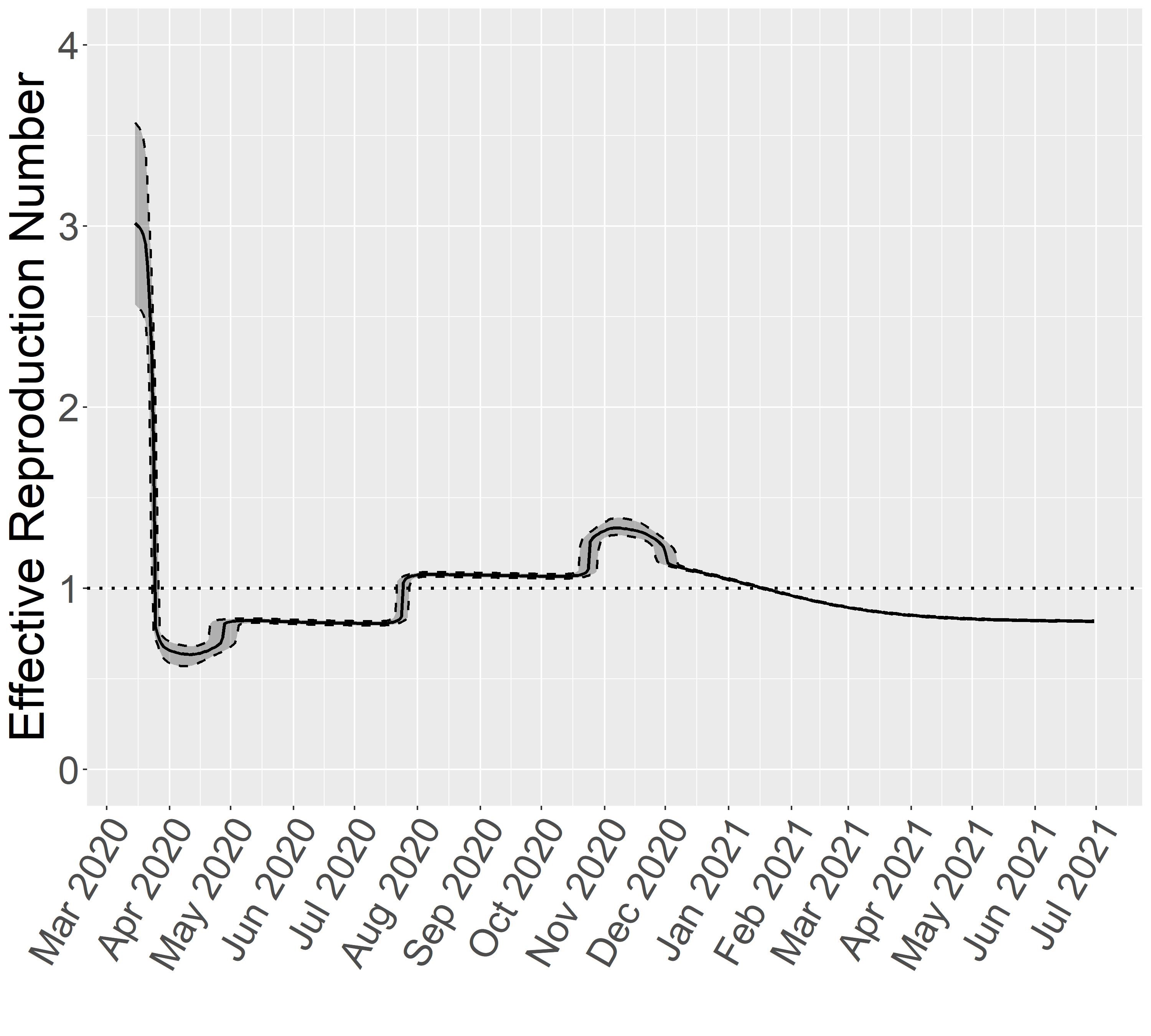}
      \caption{New York state}
    \end{subfigure}
    \begin{subfigure}{0.5\linewidth}
      \centering
      \includegraphics[width=1\linewidth]{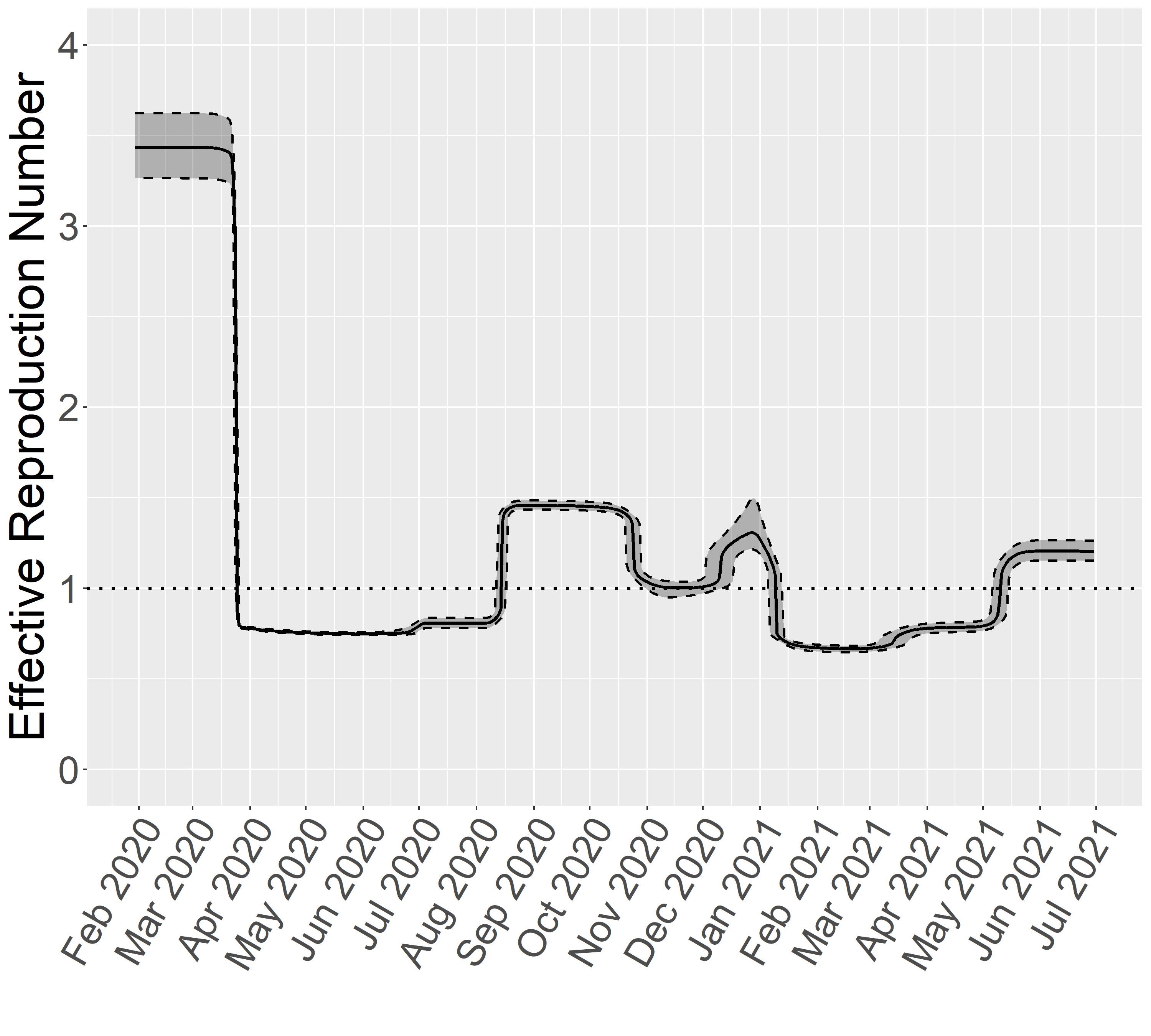}
      \caption{The United Kingdom}
    \end{subfigure}%
    \begin{subfigure}{0.5\linewidth}
      \centering
      \includegraphics[width=1\linewidth]{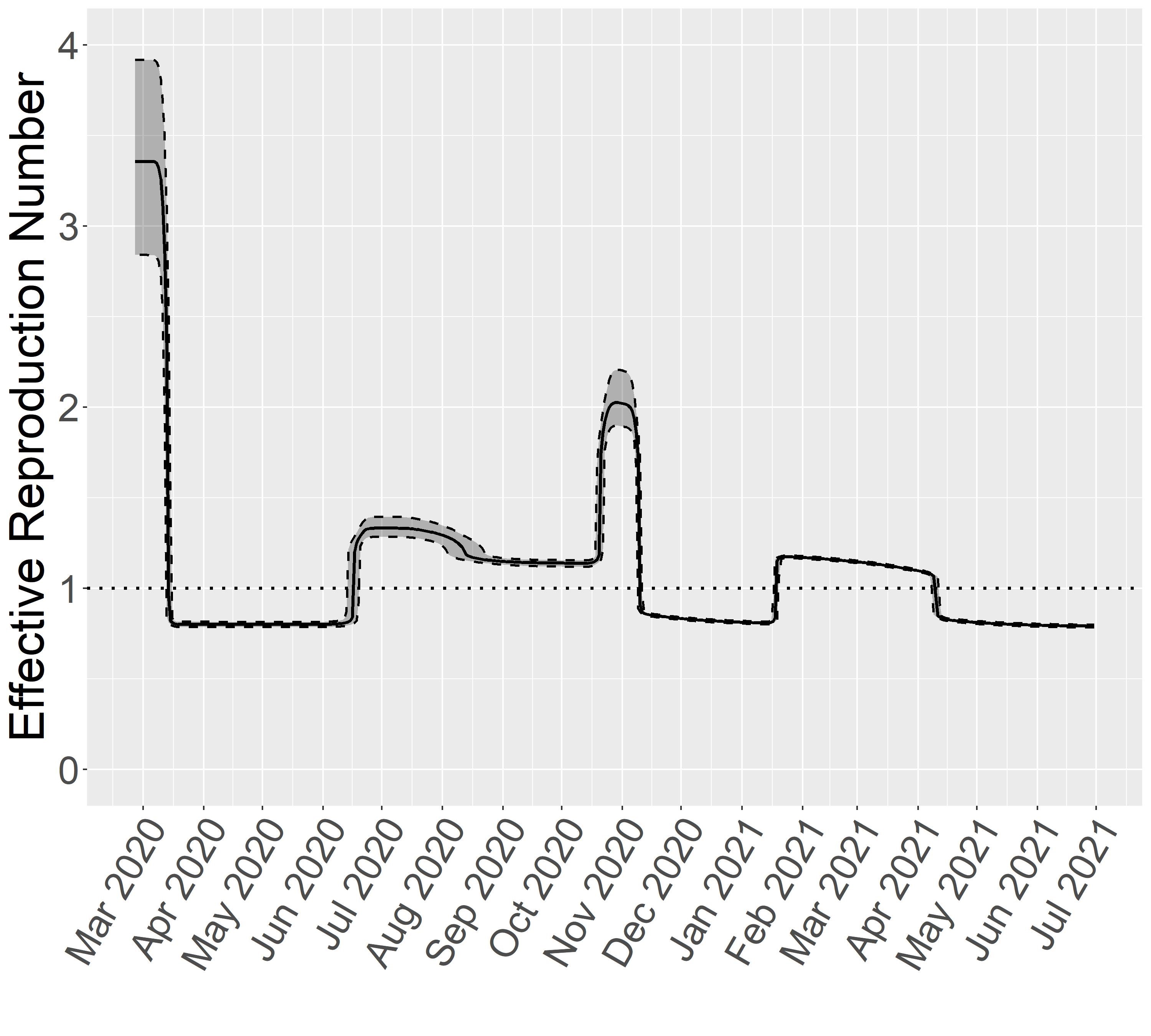}
      \caption{Greece}
    \end{subfigure}
  \caption {Estimation of Effective Reproduction Number $R_e(t)$  with 50\% Cr.I. (solid and dashed lines) based on observing deaths, fixed number of phases model.\label{fig:fixed_realdata_rt}}
\end{figure*}

\begin{figure*}[hbt!]
    \centering
    \begin{subfigure}{0.5\linewidth}
      \centering
      \includegraphics[width=1\linewidth]{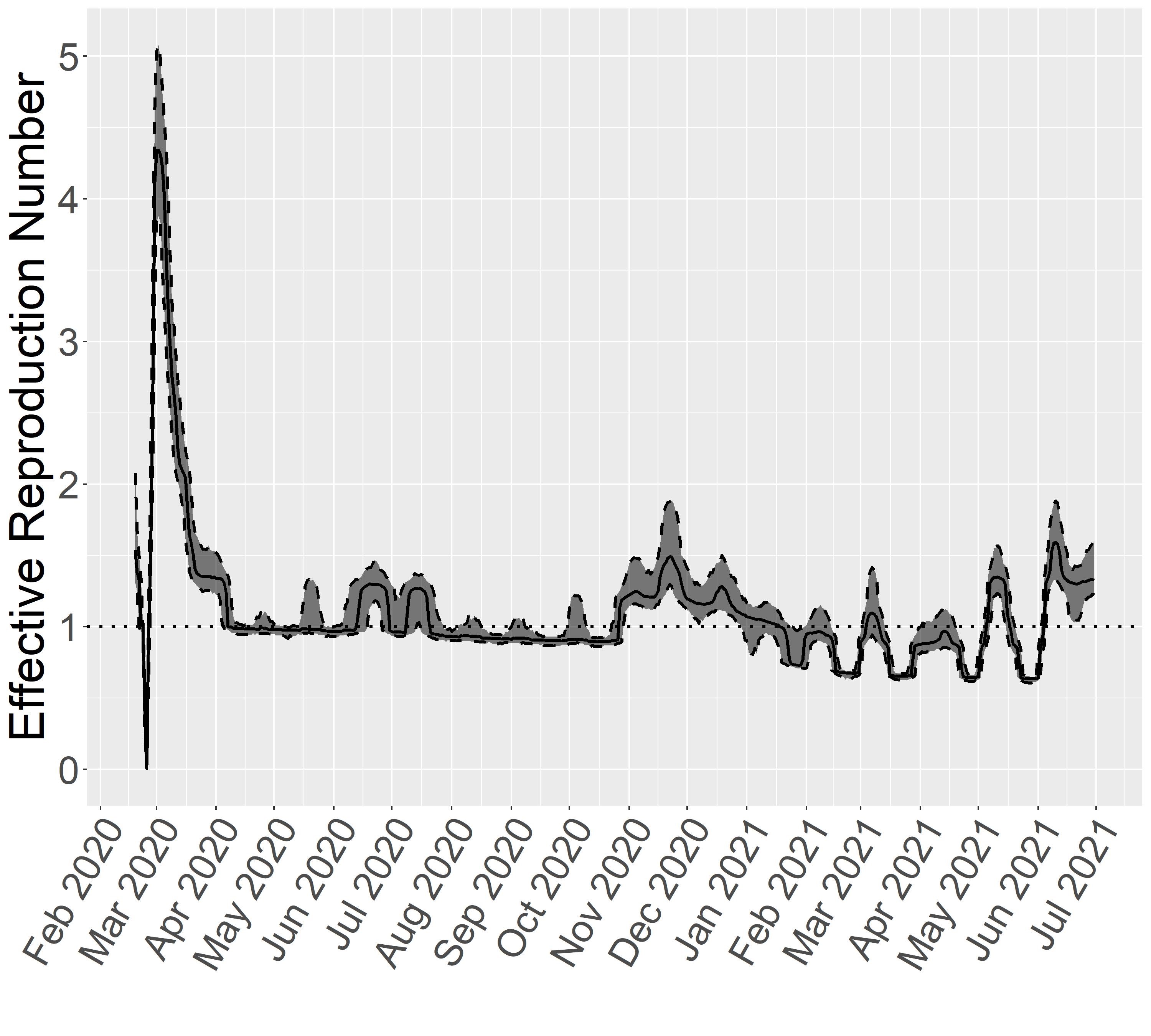}
      \caption{California state - DP model}
    \end{subfigure}%
    \begin{subfigure}{0.5\linewidth}
      \centering
      \includegraphics[width=1\linewidth]{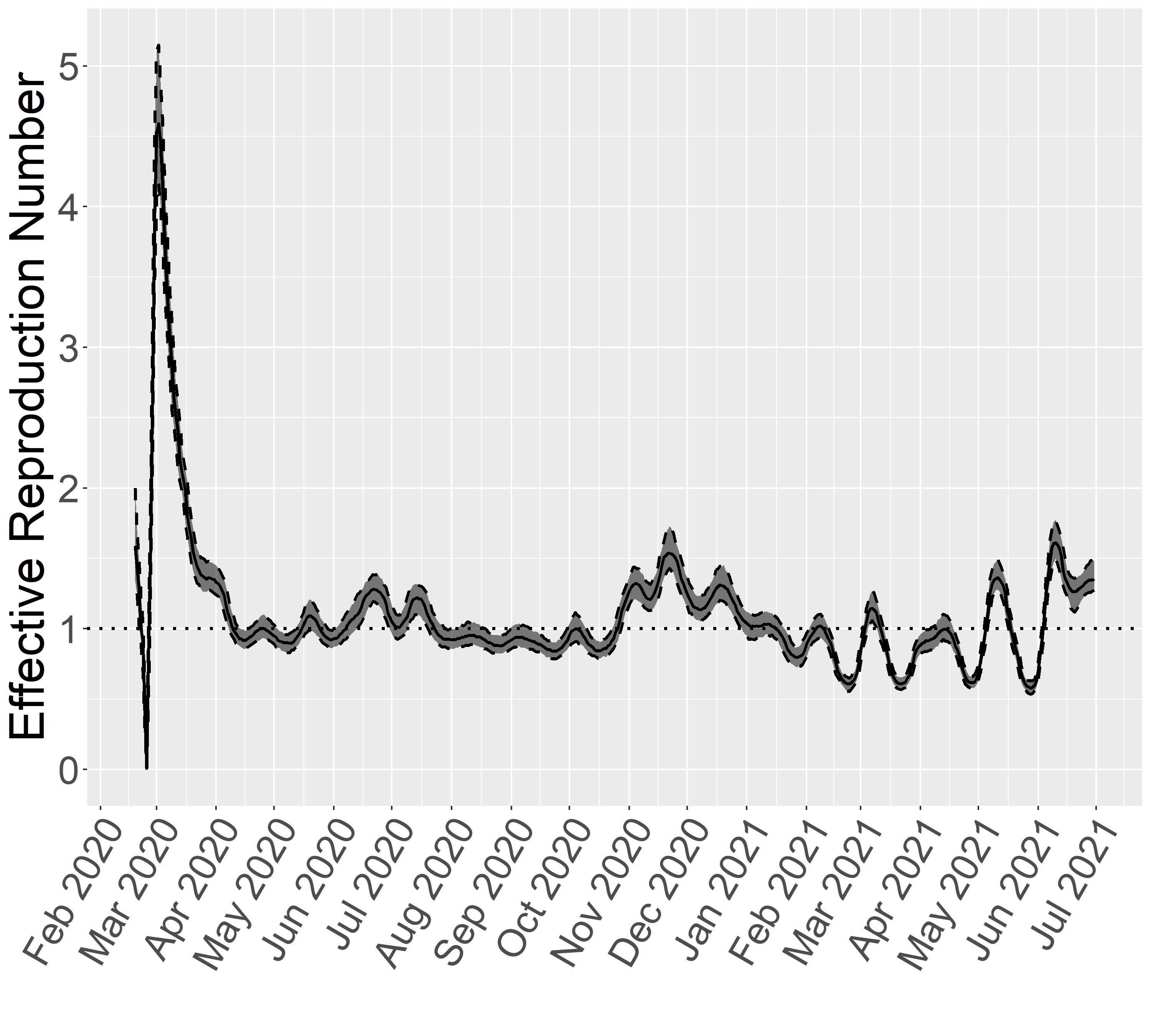}
      \caption{California state - PP model}
    \end{subfigure}
    \begin{subfigure}{0.5\linewidth}
      \centering
      \includegraphics[width=1\linewidth]{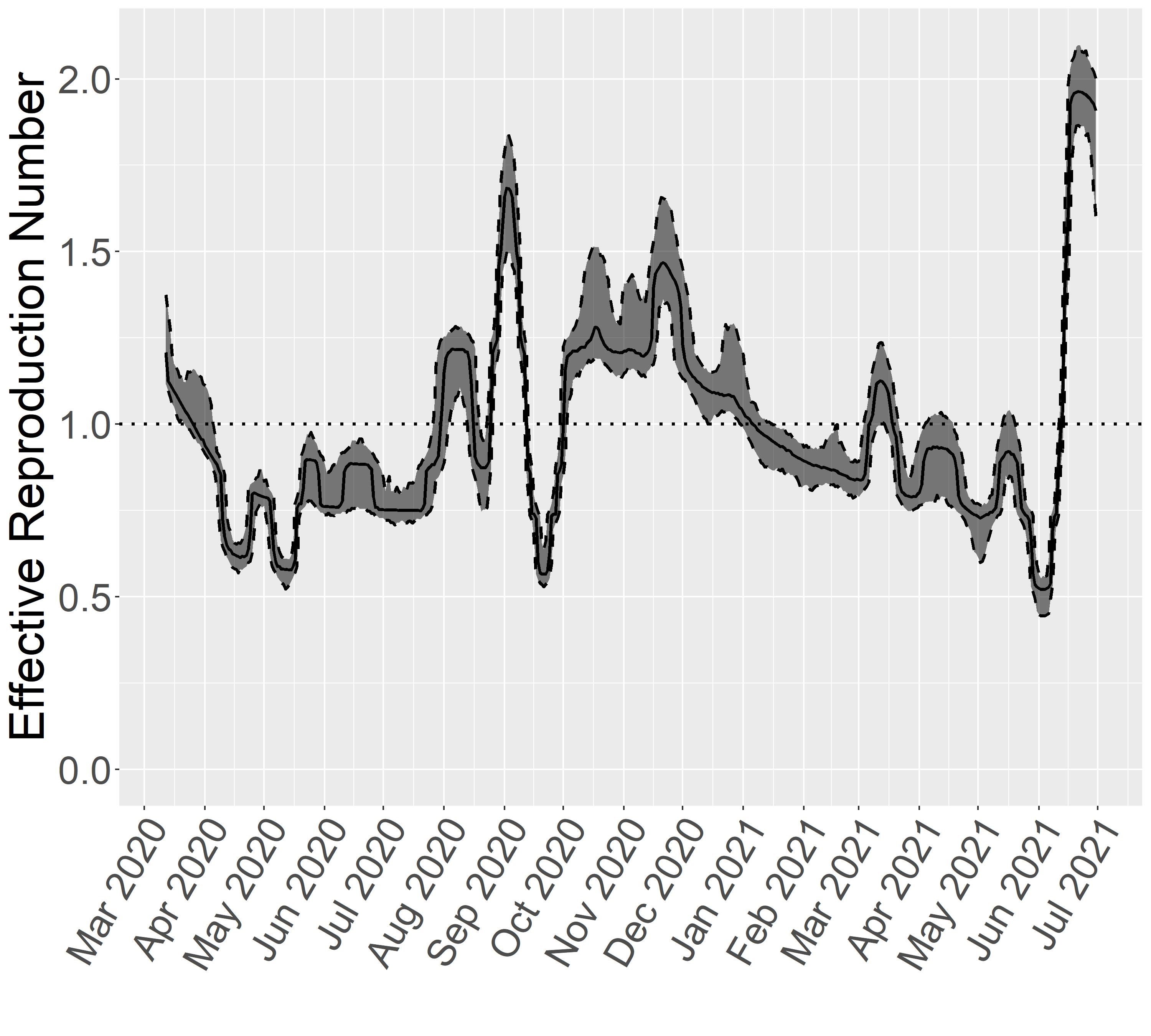}
      \caption{New York state - DP model}
    \end{subfigure}%
    \begin{subfigure}{0.5\linewidth}
      \centering
      \includegraphics[width=1\linewidth]{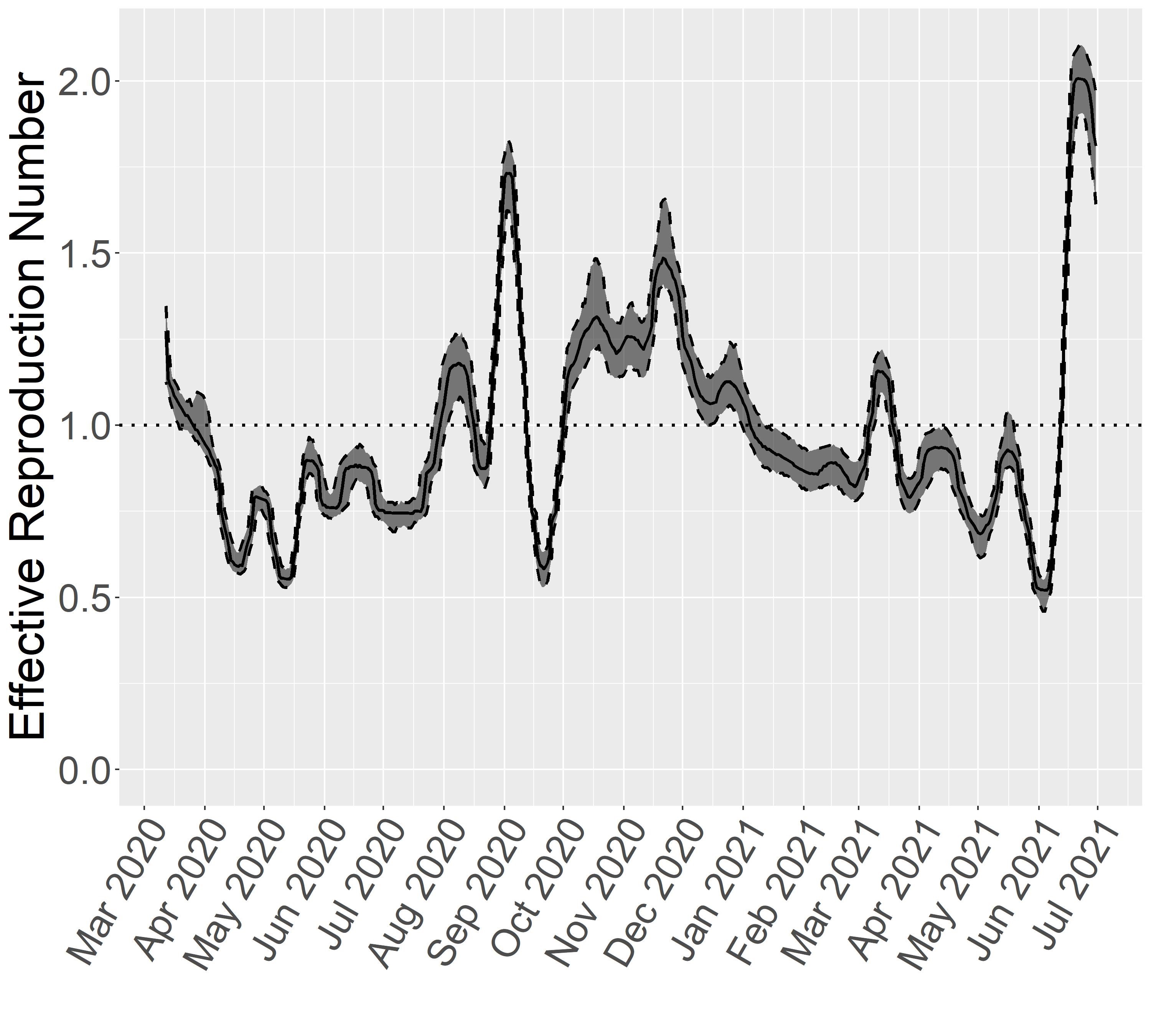}
      \caption{New York state - PP}
    \end{subfigure}
  \caption {Estimation of Effective Reproduction Number $R_e(t)$ with 95\% Cr.I. (solid and dashed lines) based on observing deaths, multi-stage approach. \label{fig:DP_PP_CA_NY_rt}}
\end{figure*}

\emph{California} was one of the first US states to report cases on the 26th of January, 2020. A state of emergency was declared on March 4, 2020, and mass/social gatherings were banned while a mandatory statewide stay-at-home order was issued on March 19, 2020. We fitted the model to daily deaths and using WAIC/LOO selected 6 changepoints. Figures \ref{fig:fixed_realdata_rt} and \ref{fig:DP_PP_CA_NY_rt} suggest that $R_e(t)$ was reduced after imposing restrictions and fell below the critical value of 1 after April 2020 when school closure was decided for the remainder of the 2019–2020 academic year. The epidemic remained under control until the summer of 2020 when $R_e(t)$ jumped slightly above 1 following a gradual relaxation of measures. On August 31, 2020, a new set of measures called `Blueprint for a Safer Economy' was applied and all models show that they were effective, alongside the gained immunity of the population, at reducing the effective reproduction number below one and keeping the epidemic under control until the first half of October 2020. All models estimate a sharp increase in $R_e(t)$, which resulted in an increase in the daily reported cases and deaths between November 2020 and January 2021. Nighttime curfew and regional stay-at-home orders were announced at the start of December 2020 whence $R_e(t)$ remained stable and began declining. The initiation of the vaccination program on early 2021 brought the epidemic under control with $R_e(t)$ remaining below 1. 

\emph{New York} state had, by April 10 2020, more confirmed cases than any country outside the US and was heavily affected at the start of the pandemic, with daily recorded deaths reaching a thousand in April. On March 15 all New York City schools were closed and on March 20 state-wide stay-at-home order was declared. As a result, the models show a drop of $R_e(t)$ below 1 from mid-March 2020 until August 2020 (Figures \ref{fig:fixed_realdata_rt} and \ref{fig:DP_PP_CA_NY_rt}). The best-performing models based on WAIC and LOO had 7 changepoints (8 distinct phases). This model estimates that after the summer of 2020, $R_e(t)$ remained above 1 up until the start of 2021 with a small increase during November and the holiday season. The DP and PP models show similar estimates for $R_e(t)$ (Figure \ref{fig:DP_PP_CA_NY_rt}). 

For the \emph{United Kingdom} a model with 8 changepoints was selected by WAIC and LOO. Until early March 2020, when a lockdown was imposed we estimate that $R_t \approx 3.5$ (Figure \ref{fig:fixed_realdata_rt}). These measures were lifted in early June and during the lockdown $R_e(t)$ remained below 1, and therefore under control. After the summer $R_e(t)$ increased above 1 and the so-called rule of six was imposed while on November 5, 2020, the second lockdown was announced. The number of reported deaths was reduced after the initiation of the vaccination program on January 4 2021. Virtually identical estimates for the UK $R_e(t)$ are inferred by the DP and PP models (Figures and additional details in the supplementary material).

\begin{figure*}[hbt!]
    \centering
      \includegraphics[width=1\linewidth]{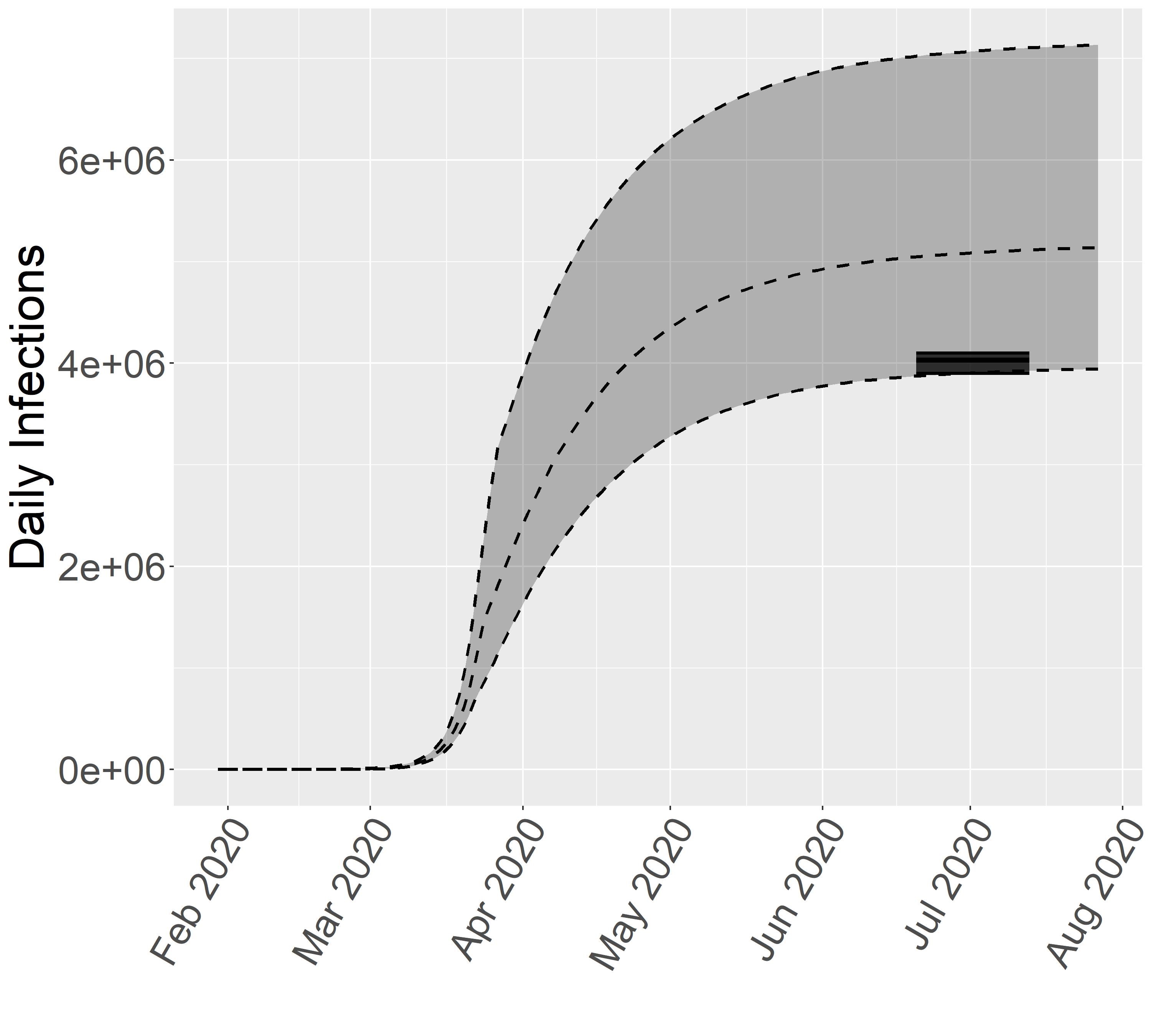}
  \caption { Cumulative sum of estimated daily infections with 95\% Cr.I. (dashed lines) and the estimation of REACT-2 with 95\% C.I. (solid lines) for the United Kingdom \label{fig:uk_cumsum_cases}}
\end{figure*}

\begin{figure*}[hbt!]
    \centering
    \begin{subfigure}{0.5\linewidth}
      \centering
      \includegraphics[width=1\linewidth]{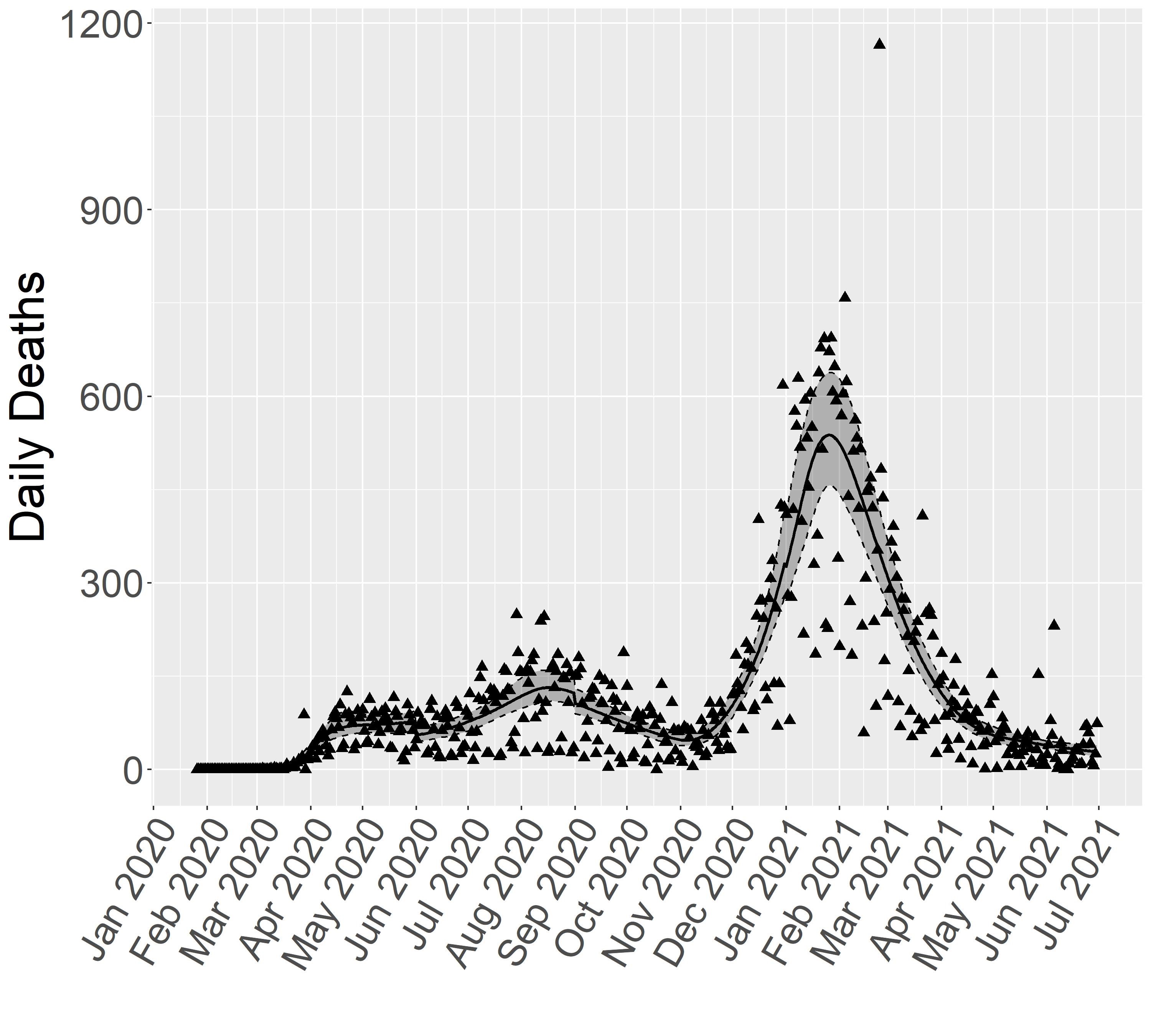}
      \caption{California state}
    \end{subfigure}%
    \begin{subfigure}{0.5\linewidth}
      \centering
      \includegraphics[width=1\linewidth]{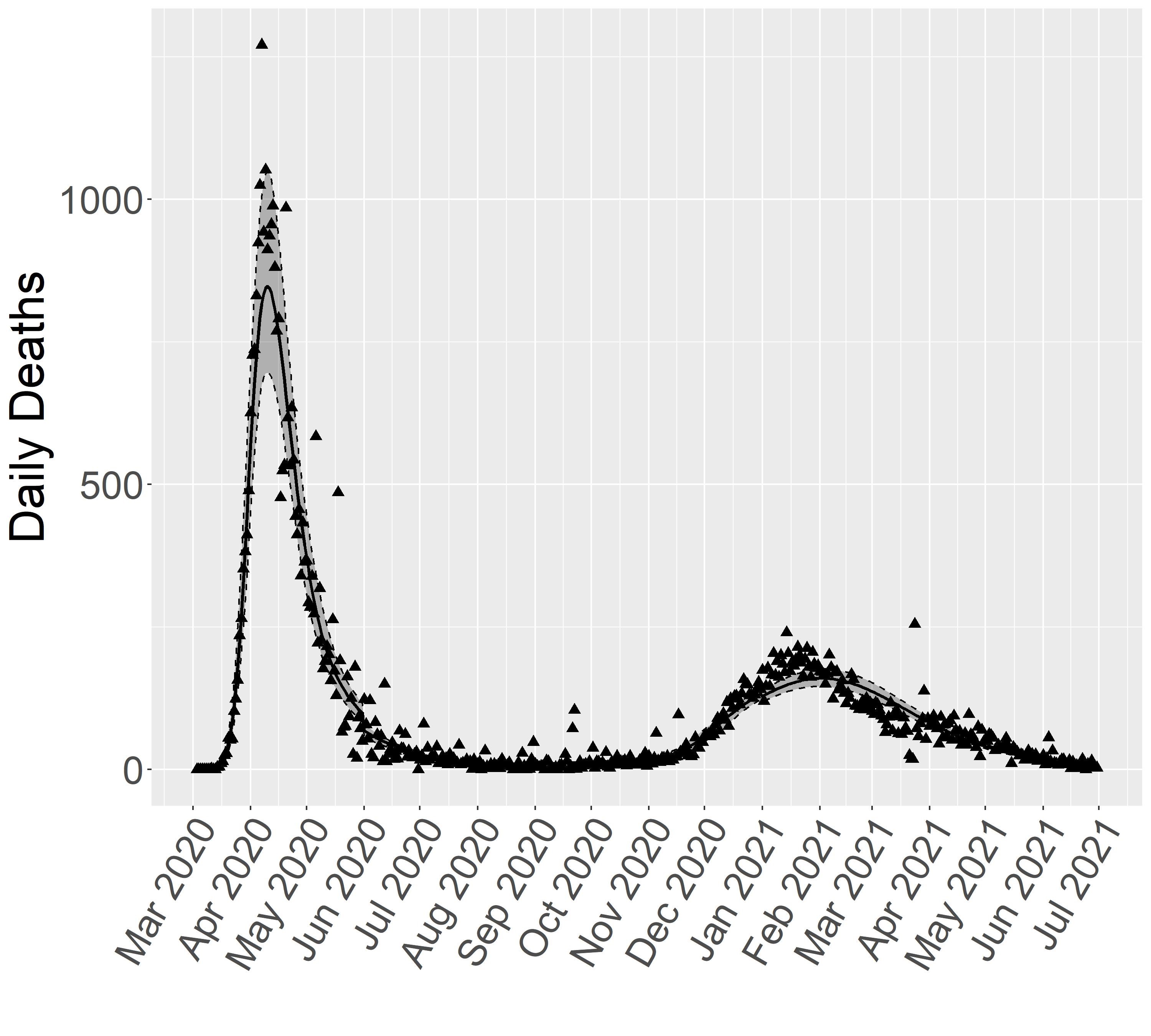}
      \caption{New York state}
    \end{subfigure}
    \begin{subfigure}{0.5\linewidth}
      \centering
      \includegraphics[width=1\linewidth]{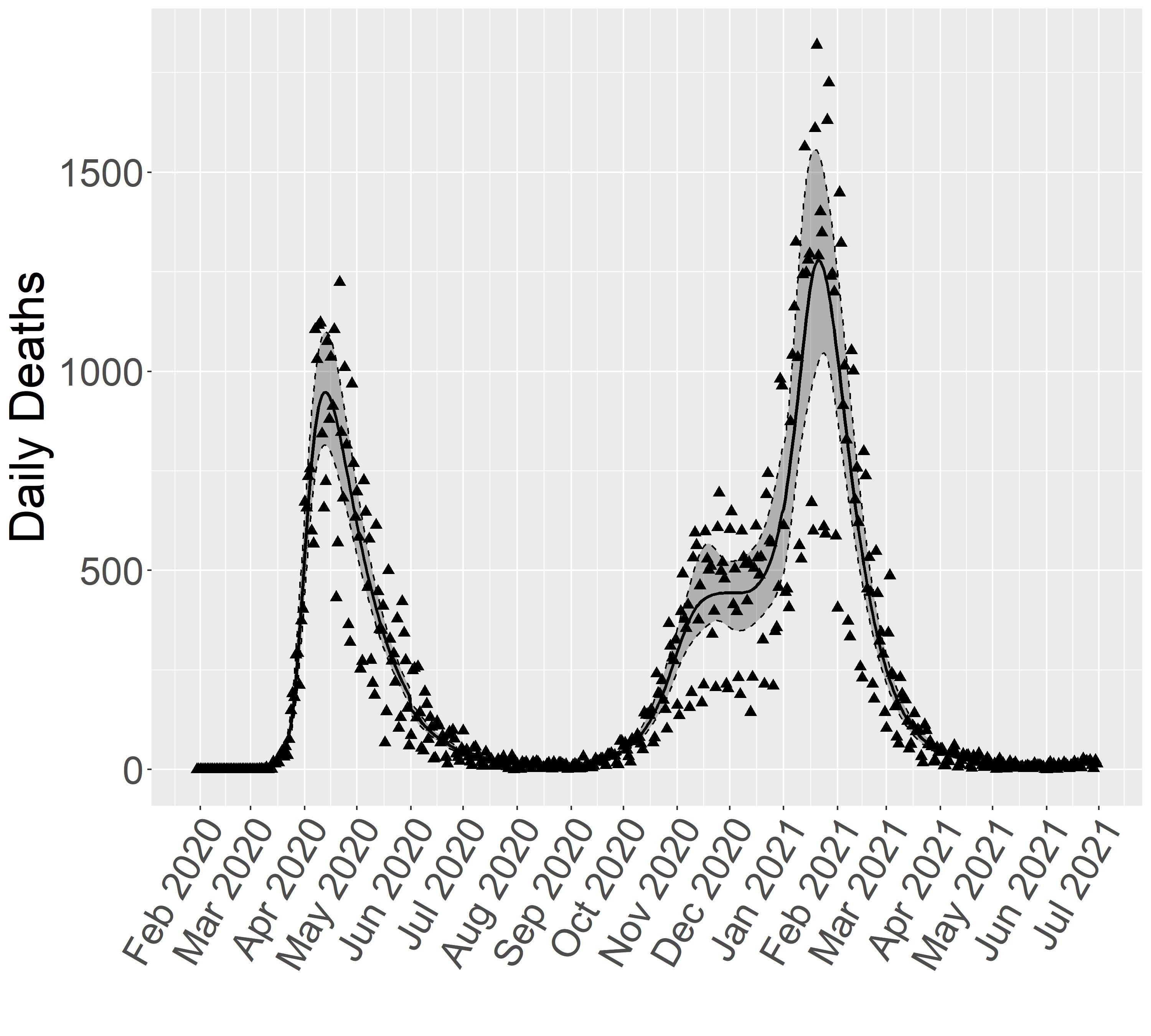}
      \caption{The United Kingdom}
    \end{subfigure}%
    \begin{subfigure}{0.5\linewidth}
      \centering
      \includegraphics[width=1\linewidth]{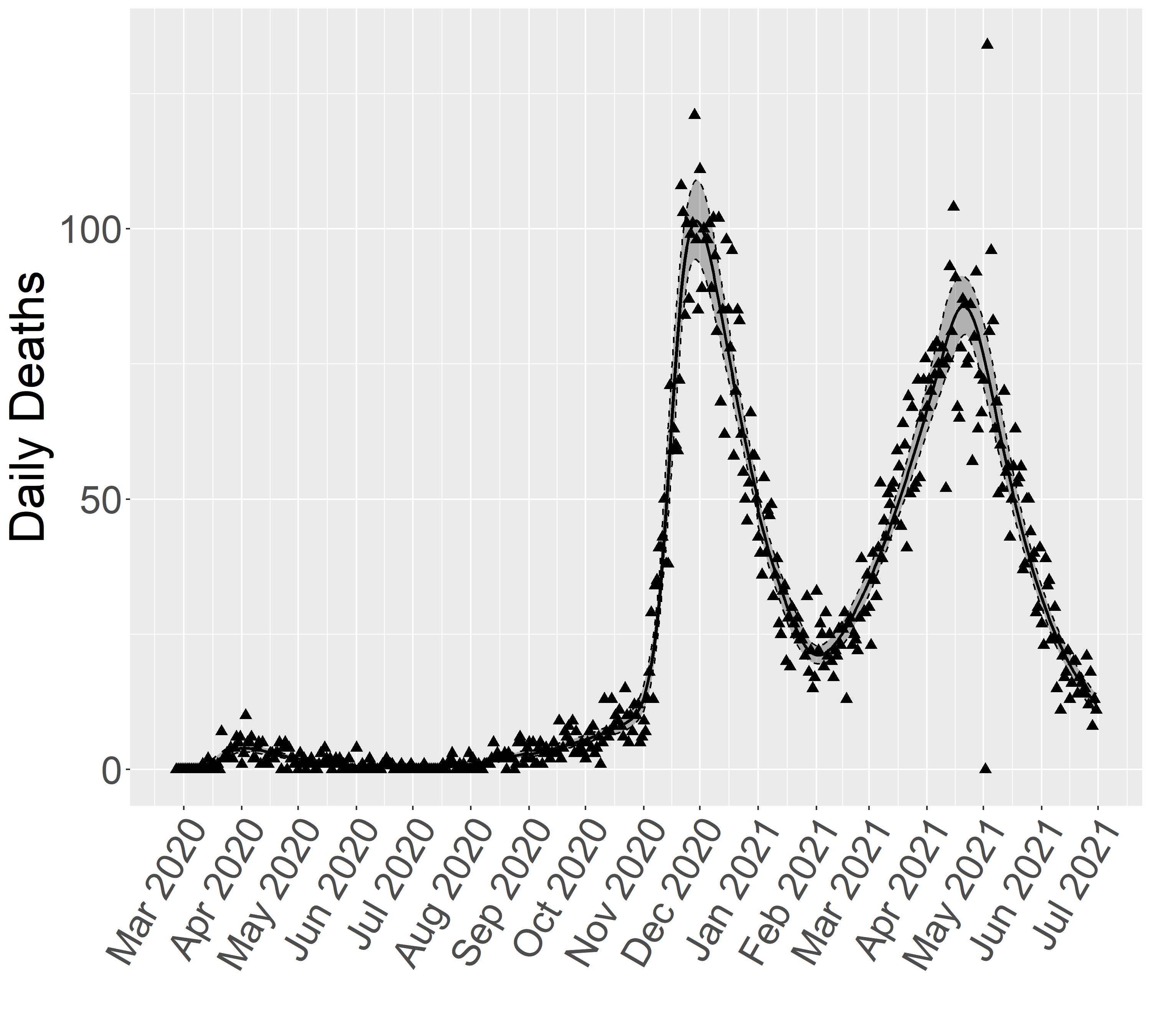}
      \caption{Greece}
    \end{subfigure}
  \caption {Reported (triangles) and estimated deaths with 50\% Cr.I. (solid and dashed lines) based on observing deaths, fixed number of phases model.\label{fig:fixed_realdata_deaths}}
\end{figure*}

We conducted an independent (or `external') validation of the model performance based upon REACT-2, an antibody prevalence study conducted in the UK with the participation of more than 100000 adults \citep{Ward2021}. This is a unique opportunity as it took place on early July 2020 when waning immunity was unlikely and provides a reasonable estimate of the total disease burden up to that time. The estimated prevalence for the adult population (children were excluded) was 6.0\% (95\% CI: 5.8, 6.1) and our estimate for the whole population is 7.5\% (95\% Cr.I.: 5.7, 10.) (Figure \ref{fig:uk_cumsum_cases}) well compatible with that independent estimate. 

For \emph{Greece} WAIC and LOO selected the 7-changepoint model. At the starting phase, we estimate $R_e(t)= 3.36$ $(sd=0.88)$ and a decrease below 1 on the first half of March 2020 (Figure \ref{fig:fixed_realdata_rt}). On March 10 the government suspended most activities, including educational, shopping and recreational while a week later all nonessential movement was restricted. The $R_e(t)$ estimate remained below 1 until early June 2020 when it increased following the lifting of restrictions. During summer $R_e(t)$ remained over 1 until November 2020 since a case spike on October led to new measures. Similar estimates for the $R_e(t)$ are obtained by the DP and PP models (Supplementary material).

The computation time was similar for the PP and DP models with the DP being faster. More importantly, we get valuable insights on the effectiveness of the measures imposed by the governments. For New York and the UK it appears that the NPIs predate the reductions in transmissibility. California and Greece adopted the measures before a large first wave, like other EU countries and US states. All regions were similar when these measures were relaxed: multiple epidemic waves emerged and the estimated $R_e(t)$ remained above 1.  

The results of our simulation experiments corroborate the findings of the application to real data from different areas. The time-ordering of the data facilitates avoiding label-switching problems typically encountered when fitting mixture models. By selecting the number of phases we capture  mortality changes in all the real-world examples (Figure \ref{fig:fixed_realdata_deaths}). The DP and PP models can infer a slightly higher number of phases but the conclusions are not materially affected. This observation is in line with \citet{Rousseau2011} who show a generally stable behaviour of such so-called overfitted mixture models, theoretically verifying the robust behaviour of the developed models.

\section{Discussion}
\label{sec:Disc}

In this article, we propose 3 models for the transmission mechanism of infectious diseases with multiple epidemic phases. We use freely available data to estimate the points in time when transmissibility changes and the realised magnitude of tho NPI effects. We adopt this approach since many of these interventions coexist or overlap and identifiability issues can arise when disentangling individual effects and the associated time lags. Essentially, one may retrospectively assess the effect of the NPIs by comparing the changes in the reproduction number with the dates that these measures were imposed. Selecting the number of phases requires multiple runs and the computation time can be an issue when nowcasting is essential for decision-making. Estimating model complexity via the DP and PP models represents an alternative approach that is computationally efficient and statistically robust.  

The DP and PP models can estimate more epidemic phases and this issue is discussed in detail in \citet{Rousseau2011} and \citet*{Miller2013-wh}. In our setting, this effect essentially relates to the start and end of the epidemic and the inherent challenges of limited information. At the start of the epidemic, such uncertainty dictates that estimates should be interpreted with caution. In the end, this is less of an issue and is mostly due to the time lag between cases and deaths. When one is working with the observed infections these issues are largely removed and inference is typically accurate throughout the duration of the data as indicated by our simulation experiments. 

The models developed in this work are assuming a homogeneous and homogeneously mixing population, like most of the work studying SARS-CoV2 transmission. This may be appropriate for large populations such as working at the state or country level since functional central limit theorems can reasonably be thought of as applicable \citep*[e.g.,][]{Andersson2000-kf}. Our models can naturally be extended when more detailed information is available and this is the subject of current research.

\section*{Acknowledgments}

This article is part of the first author's doctoral thesis, co-financed by Greece and the European Union (European Social Fund-ESF) through the Operational Programme `Human Resources Development, Education and Lifelong Learning' in the context of the Act `Enhancing Human Resources Research Potential by undertaking a Doctoral Research' Sub-action 2: IKY Scholarship Programme for PhD candidates in the Greek Universities. 

The authors are grateful to Kostas Kalogeropoulos and Petros Dellaportas for useful comments on an earlier version of this article.

\bibliographystyle{apalike}  
\bibliography{references}

\end{document}